# Effect of interface on epitaxy and magnetism in h-$RFeO_3$/$Fe_3O_4$/$Al_2O_3$ films (R=Lu, Yb)

Xiaozhe Zhang[1,2], Yuewei Yin[2], Sen Yang[1], Zhimao Yang[1*], Xiaoshan Xu[2,3*]

[1] School of Science, MOE Key Laboratory for Nonequilibrium Synthesis and Modulation of Condensed Matter, Xi'an Jiaotong University, Xi'an 710049, China

[2] Department of Physics and Astronomy, University of Nebraska, Lincoln, NE 68588, USA

[3] Nebraska Center for Materials and Nanoscience, University of Nebraska, Lincoln, NE 68588, USA

*To whom all correspondence should be addressed.

zmyang@xjtu.edu.cn, xiaoshan.xu@unl.edu

**Abstract**

We have carried out the growth of h-RFeO$_3$ (001) (R=Lu, Yb) thin films on Fe$_3$O$_4$ (111)/Al$_2$O$_3$ (001) substrates, and studied the effect of the h-RFeO$_3$ (001)/Fe$_3$O$_4$ (111) interfaces on the epitaxy and magnetism. The observed epitaxial relations between h-RFeO$_3$ and Fe$_3$O$_4$ indicates an unusual matching of Fe sub-lattices rather than a matching of O sub-lattices. The out-of-plane direction was found to be the easy magnetic axis for h-YbFeO$_3$ (001) but the hard axis for Fe$_3$O$_4$ (111) in the h-YbFeO$_3$ (001)/Fe$_3$O$_4$ (111)/Al$_2$O$_3$ (001) films, suggesting a perpendicular magnetic alignment at the h-YbFeO$_3$ (001)/Fe$_3$O$_4$ (111) interface. These results indicate that Fe$_3$O$_4$ (111)/Al$_2$O$_3$ (001) could be a promising substrate for epitaxial growth of h-RFeO$_3$ films of well-defined interface and for exploiting their spintronic properties.

**Introduction**

Thin film epitaxy and heterostructures have been shown effective in exploiting the rich properties in transition metal oxides, [1–3] taking advantage of their structural sensitivity and the complex electronic structures at the interface. While this is encouragingly true for the perovskite family of cubic or distorted cubic symmetries [4–6], the shortage of structurally compatible substrates and well-defined interfaces hinders the study of other families of materials (e.g. of trigonal or hexagonal symmetry) using thin film epitaxy.

Here we attack the problem of thin film epitaxy of hexagonal ferrites. Hexagonal ferrites h-$RFeO_3$ (R=Lu, Yb) simultaneous exhibit ferroelectric and weakly ferromagnetic orders; [7–10] they belong to a class of complex materials called multiferroics which are promising in compact and energy efficient information storage and processing. [11,12] The few choices of substrates for preparing thin films of h-$RFeO_3$ include $Al_2O_3$ (001), yttrium stabilized zirconia or YSZ (111), and Pt (111). [8–10,13–15] A significant number of defects are expected at the interfaces between these substrates and films due to the larger lattice mismatch (> 5%), which undermines the study of the intrinsic properties of the thin films and interfaces, as well as the fabrication of devices.

We have recently demonstrated epitaxial growth of $Fe_3O_4$ (111) on $Al_2O_3$ (001) with high crystallinity and smooth surface of atomic terraces. [16] The in-plane lattice constant of $Fe_3O_4$ (111) is 11.87 Å ($\sqrt{2}$ times the lattice constant), [16] which matches twice of that of h-$LuFeO_3$ and h-$YbFeO_3$ in the basal plane (5.96 Å and 5.99 Å respectively) [7,17] within a 1% differences. In addition, previous growth of h-$LuFeO_3$ indicates that $Fe_3O_4$ (111) layers may naturally occur in the h-LuFeO$_3$ films, under the reducing environments (oxygen deficient or Fe rich). [15,18] Therefore, the Fe$_3$O$_4$ (111)/Al$_2$O$_3$ (001) may be a compatible substrate for h-RFeO$_3$, with well-defined interfaces due to the small lattice mismatch. In addition, the Fe$_3$O$_4$ (111) layer can be employed as a bottom electrode for studying the effect of electric field in h-RFeO$_3$. Therefore, it is intriguing to prepare the h-RFeO$_3$ films on Fe$_3$O$_4$ (111)/Al$_2$O$_3$ (001) and study their properties, especially multiferroicity. As a foundation of these studies, the intrinsic properties at the h-RFeO$_3$/Fe$_3$O$_4$ interface and their effect on the epitaxy and magnetism of the films are of great importance.

We have prepared h-RFeO$_3$ (001)/Fe$_3$O$_4$ (111)/Al$_2$O$_3$ (001) films using pulsed laser deposition. The structural characterizations show that the films are epitaxial and the lattices of h-RFeO$_3$ (001) and Fe$_3$O$_4$ (111) do align according to their in-plane lattices. The interface appears to be critical in the epitaxial relations and the magnetic alignment between h-RFeO$_3$ and Fe$_3$O$_4$. These results demonstrate that the Fe$_3$O$_4$ (111)/Al$_2$O$_3$ (001) is a promising substrate for preparing hexagonal ferrites thin films with well-defined film-substrate interfaces.

**Experimental**

The h-RFeO$_3$ (001)/Fe$_3$O$_4$ (111)/Al$_2$O$_3$ (001) films were grown using pulsed laser depositions. [16,18] The Fe$_3$O$_4$ (111) thin films (5-30 nm) were deposited epitaxially on Al$_2$O$_3$ (001) substrates, as described in our previous work. [16] The h-RFeO$_3$ thin films (5-30 nm) were deposited epitaxially on top of the Fe$_3$O$_4$ (111) thin films, in a 5 mTorr Ar environment at 750 °C with a laser fluence of ~1 J cm$^{-2}$, and a repetition rate of 2 Hz. [9,14,19,20] The epitaxial relations between different layers in the films were studied with in-situ reflection high energy electron diffraction (RHEED) and ex-situ X-ray diffractions (XRD). The θ-2θ scans of X-ray diffraction were carried out using a Rigaku D/Max-B diffractometer, with a cobalt K-$\alpha$ source ($\lambda = 1.79$ Å). The rocking curve (ω scan), φ scan, and reciprocal space mapping were studied using a Rigaku Smartlab diffractometer, with a copper K-$\alpha$ source ($\lambda = 1.54$ Å). The surface morphology of the films was studied using the atomic force microscopy (AFM) with a Bruker Dimension ICON. The magnetic properties of the h-RFeO$_3$ (001)/Fe$_3$O$_4$ (111)/Al$_2$O$_3$ (001) films were studied using a superconducting quantum interference device (SQUID) magnetometer.

**Results and Discussion**

First, we investigate the epitaxial relations of the h-RFeO$_3$/Fe$_3$O$_4$/Al$_2$O$_3$ films using structural characterization. Fig. 1(a) shows the larger range θ-2θ scan of XRD of the h-RFeO$_3$ (001)/Fe$_3$O$_4$ (111)/Al$_2$O$_3$ (001) films; no impurity phase is observed. The small-range scans were taken around the h-RFeO$_3$ (002) and Fe$_3$O$_4$ (111) peaks [Fig. 1(b)]. The Laue oscillations indicate that these films have flat surfaces, [7] which is consistent with the surface roughness (< 1 nm) demonstrated by AFM (see supplementary materials [21]). The φ scans of these films indicate that the Fe$_3$O$_4$ (111) and h-RFeO$_3$ (001) layers are indeed epitaxial on the Al$_2$O$_3$ (001) substrates (see supplementary materials [21]). Lattice constants of the h-RFeO$_3$ layers were measured using reciprocal space mapping (RSM) (see supplementary materials [21]); the results show that for h-LuFeO$_3$, $a$ = 5.963 Å, $c$ = 11.92 Å and for h-YbFeO$_3$, $a$ = 6.021 Å, $c$ = 12.07 Å, in agreement with the previous measurements. [7,17]

The RHEED patterns obtained on different layers reveal their epitaxial relations. As shown in Fig. 2, two directions of incident electron beams that are perpendicular to each other ($Al_2O_3$ <120> and $Al_2O_3$ <100>) were used. The RHEED patterns of all the layers ($Al_2O_3$, $Fe_3O_4$, h-$LuFeO_3$, and h-$YbFeO_3$) are in accord with in-plane triangular lattices, suggesting a relation $Al_2O_3$ (001)//$Fe_3O_4$ (111)//h-$RFeO_3$ (001). Using the lattice constant of $Al_2O_3$ as the calibration, one can estimate the lattice constants of the epilayers: $a$ = 8.31±0.08 Å for $Fe_3O_4$, $a$ = 5.92±0.06 Å for h-$LuFeO_3$, and $a$ = 6.02±0.06 Å for h-$YbFeO_3$, in line with values found in the XRD measurements.

The in-plane epitaxial relation $Al_2O_3$ <100>//$Fe_3O_4$ <-211>//h-$RFeO_3$ <1-10> can be extracted from the RHEED pattern (Fig. 2), as well as from the XRD φ scan (see supplementary materials [21]). Previously, it was found that when h-$LuFeO_3$ was deposited directed on the $Al_2O_3$ (001) substrates, the in-plane epitaxial relation was $Al_2O_3$ <001> // h-$LuFeO_3$ <001>, which is different from the relation found in the h-$RFeO_3$ (001)/$Fe_3O_4$ (111)/$Al_2O_3$ (001) films here. Obviously, this difference comes from the peculiar in-plane epitaxial relation between $Fe_3O_4$ (111) and $Al_2O_3$ (001) layers, and that between $Fe_3O_4$ (111) and h-$RFeO_3$ (001) layers.

As discussed in our previous work [16], the epitaxial relation $Al_2O_3$ <100> // $Fe_3O_4$ <-211> comes from the matching of the in-plane oxygen sub-lattice. [10,16] In this case, the lattice constants of the in-plane oxygen triangular sub-lattices are approximately 2.92 Å and 2.85 Å for $Al_2O_3$ (100) and $Fe_3O_4$ (111) respectively, [22,23] which means a modest 2.5% mismatch. Since there is a 30° rotation between the *a*-axis of $Al_2O_3$ and that of the triangular oxygen sub-lattice in the basal plane $Al_2O_3$ (001), to share the oxygen layer, the angle between the in-plane *a*-axis of $Fe_3O_4$ (111) ($Fe_3O_4$ <0-11>), and that of the $Al_2O_3$ ($Al_2O_3$ <100>) is expected to be 30° (or 90° considering the six-fold rotational symmetry), which is observed in Fig. 2 (b) and

(c).

On the other hand, the matching of the oxygen sub-lattice between $Fe_3O_4$ (111) and h-$RFeO_3$ (001) is more complex, because the lattice constants of the in-plane oxygen triangular sub-lattices in h-$RFeO_3$ are approximately 3.45 Å, which is about 20% larger than that of the $Al_2O_3$ (001). Nevertheless, matching the Fe sub-lattices between $Fe_3O_4$ (111) and h-$RFeO_3$ (001) appears to be reasonable. As shown in the side view of $Fe_3O_4$ (111) plane [Fig. 3(a)], there are two kinds of Fe layers that are parallel to the $Fe_3O_4$ (111) plane (Fig. 3(b) and (c)). For one of the layers in $Fe_3O_4$ (111) that is shown in Fig. 3(c), the in-plane lattice constant is 3.43 Å, [23] which matches the in-plane Fe sub-lattice constant 3.44-3.45 Å in h-$RFeO_3$ [Fig 3(d) and(e)] [7,17] with a less than 1% difference. Therefore, the $Fe_3O_4$ (111) and h-$RFeO_3$ (001) could share the Fe sub-lattice on the interface, which leads to an epitaxial relation $Fe_3O_4$ <01-1>//h-$RFeO_3$ <100>, as shown in Fig. 2(c), 2(e), and 2(g).

The RHEED images in Fig. 2(e) and 2(g) show strong streaks separated by two weaker streaks, a patterns that is typical with a structural distortion with a propagation vector (1/3,1/3,0). [9] In the case of h-$RFeO_3$, this structural distortion is the rotation of the $FeO_5$ trigonal bipyramid and the buckling of the $LuO_2$ layer ($K_3$ mode), which induces the displacements of the atoms along the $c$-axis ($\Gamma_2^-$ mode), the ferroelectricity, and the canting of magnetic moments on Fe [9,19,24–26]. Therefore, it appears that the structural distortion that is critical for the multiferroicity in h-$RFeO_3$ is maintained in the h-$RFeO_3$/$Fe_3O_4$/$Al_2O_3$ films.

Next, we investigate the magnetic anisotropies of the h-$YbFeO_3$ (001)/$Fe_3O_4$ (111)/$Al_2O_3$ (001) films. For thin films, the magnetic anisotropy may come from the crystal structure (magneto-crystalline anisotropy) and from the dimension (shape anisotropy). The shape anisotropy is generated by the anisotropy of the depolarization factor in a film due to its quasi two-dimensional shape. While the magneto-crystalline anisotropy depends on the specific crystal structure, for the shape anisotropy of a thin film, the hard axis is always along the out-of-plane direction. In the $Fe_3O_4$ films, the magneto-crystalline anisotropy is often dominated by those created by the anti-phase boundaries. [27–29] This type of anisotropy exists for all field directions, contributes little to the remanence and coercivity, and results in unsaturated magnetization up to 70 kOe. [27–29] For the $Fe_3O_4$ films, the shape anisotropy has a much smaller energy scale. Therefore, the two types of anisotropy are manifested in different field ranges. While the shape anisotropy governs the remanence and coercivity at the low field, the high field behavior of the magnetizations of the $Fe_3O_4$ films are determined by the magneto-crystalline anisotropy created by the anti-phase boundaries (see also the supplementary materials [21]). For h-$RFeO_3$, the ferromagnetic order is parasitic to the antiferromagnetic order in which all the Fe moments lie in the basal plane. [9] The ferromagnetic magnetizations in h-$RFeO_3$ originate from the canting of the Fe moments toward the out-of-plane direction. The in-plane magnetization from Fe sites is symmetry forbidden. Therefore, in the h-$RFeO_3$ (001) films, the easy axis according to the magneto-crystalline anisotropy is along the out-of-plane direction and the shape anisotropy is not expected to play a role. In h-$LuFeO_3$, the saturation magnetization is small ($\approx$0.02 $\mu_B$/f.u.) because it only comes from the magnetic canting on the Fe sites [9,15]. In h-$YbFeO_3$, the paramagnetic Yb sites can be polarized by the exchange field of the Fe ferromagnetic magnetizations [8,10], and contribute to the total magnetizations. Due to the paramagnetic nature of the Yb sites, this contribution is large at low temperature (~1 $\mu_B$/f.u.) but drops rapidly at high temperature. [10]

Figure 4 (a) displays the magnetic hysteresis loops of a $Fe_3O_4$ (8.5 nm)/$Al_2O_3$ (001) (M1) film, an h-$YbFeO_3$ (25 nm)/$Fe_3O_4$ (11 nm)/$Al_2O_3$ (MY1) film, and an h-$YbFeO_3$ (21 nm)/$Fe_3O_4$ (20 nm)/$Al_2O_3$ (MY2) film at 10 K, with the magnetic field along the out-of-plane direction. The behavior of the M1 ($Fe_3O_4$) film is in line with a hard axis along the out of plane direction caused by the shape anisotropy (see also the supplementary materials [21]), as demonstrated by the small coercivity and magnetic remanence. [16,30]

The major features in the hysteresis loops in the h-$YbFeO_3$/$Fe_3O_4$/$Al_2O_3$ films (MY1, MY2) in Fig. 4(a) can be understood in terms of the combined magnetization of the $Fe_3O_4$ layer and the h-$YbFeO_3$ layer, assuming that their corresponding magnetic anisotropies are preserved. According to the previous work, for a h-$YbFeO_3$ (001) film in an out-of-plane field, the magnetic remanence is more than half of the magnetization at 10 kOe. [7,8] In contrast, for the $Fe_3O_4$ (111) film in an out-of-plane field, the magnetic remanence is much smaller [see Fig. 4(a)]. Therefore, for both MY1 and MY2 films in an out-of-plane field, the magnetic remanence appears to come mostly from the contribution of the h-$YbFeO_3$ layers [see Fig. 4(a)]: by adding a h-$YbFeO_3$ layer on top a $Fe_3O_4$ layer (MY1 compared with M1), the magnetic remanence increases dramatically; in contrast, increasing the thickness of the $Fe_3O_4$ layer (MY2 compared with MY1) does not affect the magnetic remanence significantly; the boost of magnetic remanence in MY1 and MY2 compared with that in M1 [Fig. 4(b)], which is obtained by adding the h-$YbFeO_3$ layer, drops dramatically at 50 K and becomes much less significant when $T \geq 100$ K, consistent with the expected disappearance of ferrimagnetism in h-$YbFeO_3$ above 120 K. [8] According to the previous work, for a h-$YbFeO_3$ (001) film in an out-of-plane field, the coercivity is in the range of 3-6 kOe, [8] which can be identified from the step-like magnetization in MY1 (on top of the background of the gradual magnetization of $Fe_3O_4$) at about 5 kOe. For the film MY2, this step in the hysteresis loop is smeared because the $Fe_3O_4$ layer is thicker than that in MY1. These results suggest that the h-$RFeO_3$/$Fe_3O_4$ interfaces comprise two magnetic materials with different anisotropy; the out-of-plane direction is an easy axis for the h-$RFeO_3$ (001) layer but a hard axis for the $Fe_3O_4$ (111) layer (see also the supplementary materials [21]). Further investigations on the magnetic interactions between the $Fe_3O_4$ and h-$YbFeO_3$ layer may benefit from the element specific method in magnetic characterizations. [31]

**Conclusion**

The epitaxial growth of h-$RFeO_3$ (001) films on $Fe_3O_4$ (111)/$Al_2O_3$ (001) substrates has been demonstrated using pulsed laser depositions. The lattice constants and the epitaxial relations between h-$RFeO_3$ (001) and $Fe_3O_4$ (111), suggests a small mismatch at the interface. The h-$RFeO_3$ (001) crystal orientation at the interface, and the shape anisotropy in $Fe_3O_4$ (111), lead to the perpendicular alignment of magnetization at the h-$RFeO_3$ (001)/$Fe_3O_4$ (111) interface, which could be interesting in exploiting spintronic applications. Furthermore, the conductive nature of $Fe_3O_4$ will be beneficial in studying the multiferroicity of h-$RFeO_3$ in an electric field, especially the voltage controlled switch of magnetizations proposed by theory. [26]

**Acknowledgement**

This project was primarily supported by the National Science Foundation (NSF), DMR under Award DMR-1454618. Z.M.Y and S.Y. acknowledge the support from the National Science

# Reference

[1]   M. E. Lines and A. M. Glass, *Principles and Applications of Ferroelectrics and Related Materials* (Clarendon Press, Oxford [Eng.], 1977).

[2]   Y. Tokura and H. Y. Hwang, Nat Mater **7**, 694 (2008).

[3]   A. P. Ramirez, J. Phys. Condens. Matter **9**, 8171 (1997).

[4]   P. Zubko, S. Gariglio, M. Gabay, P. Ghosez, and J.-M. Triscone, Annu. Rev. Condens. Matter Phys. **2**, 141 (2011).

[5]   L. W. Martin, Y. H. Chu, and R. Ramesh, Mater. Sci. Eng. R Reports **68**, 89 (2010).

[6]   N. Feature, Nature **459**, 28 (2009).

[7]   H. Iida, T. Koizumi, Y. Uesu, K. Kohn, N. Ikeda, S. Mori, R. Haumont, P.-E. Janolin, J.-M. Kiat, M. Fukunaga, and Y. Noda, J. Phys. Soc. Japan **81**, 24719 (2012).

[8]   Y. K. Jeong, J. Lee, S. Ahn, S.-W. Song, H. M. Jang, H. Choi, and J. F. Scott, J. Am. Chem. Soc. **134**, 1450 (2012).

[9]   W. Wang, J. Zhao, W. Wang, Z. Gai, N. Balke, M. Chi, H. N. Lee, W. Tian, L. Zhu, X. Cheng, D. J. Keavney, J. Yi, T. Z. Ward, P. C. Snijders, H. M. Christen, W. Wu, J. Shen, and X. Xu, Phys. Rev. Lett. **110**, 237601 (2013).

[10]  X. Xu and W. Wang, Mod. Phys. Lett. B **28**, 1430008 (2014).

[11]  N. A. Spaldin, S. W. Cheong, and R. Ramesh, Phys. Today **63**, 38 (2010).

[12]  H. Schmid, Ferroelectrics **162**, 317 (1994).

[13]  S. M. Disseler, J. A. Borchers, C. M. Brooks, J. A. Mundy, J. A. Moyer, D. A. Hillsberry, E. L. Thies, D. A. Tenne, J. Heron, M. E. Holtz, J. D. Clarkson, G. M. Stiehl, P. Schiffer, D. A. Muller, D. G. Schlom, and W. D. Ratcliff, Phys. Rev. Lett. **114**, 217602 (2015).

[14]  S. Cao, T. R. Paudel, K. Sinha, X. Jiang, W. Wang, E. Y. Tsymbal, X. Xu, and P. A. Dowben, J. Phys. Condens. Matter **27**, 175004 (2015).

[15]  J. A. Moyer, R. Misra, J. A. Mundy, C. M. Brooks, J. T. Heron, D. A. Muller, D. G. Schlom, and P. Schiffer, APL Mater. **2**, 12106 (2014).

[16]  X. Zhang, S. Yang, Z. Yang, and X. Xu, J. Appl. Phys. **120**, 85313 (2016).

[17]  E. Magome, C. Moriyoshi, Y. Kuroiwa, A. Masuno, and H. Inoue, Jpn. J. Appl. Phys. **49**, 09ME06 (2010).

[18] W. Wang, Z. Gai, M. Chi, J. D. Fowlkes, J. Yi, L. Zhu, X. Cheng, D. J. Keavney, P. C. Snijders, T. Z. Ward, J. Shen, and X. Xu, Phys. Rev. B **85**, 155411 (2012).

[19] S. Cao, X. Zhang, T. R. Paudel, K. Sinha, X. Wang, X. Jiang, W. Wang, S. Brutsche, J. Wang, P. J. Ryan, J.-W. Kim, X. Cheng, E. Y. Tsymbal, P. A. Dowben, and X. Xu, J. Phys. Condens. Matter **28**, 156001 (2016).

[20] S. Cao, X. Zhang, K. Sinha, W. Wang, J. Wang, P. A. Dowben, and X. Xu, Appl. Phys. Lett. **108**, 202903 (2016).

[21] See supplementary material at for more detailed information on X-Ray diffraction, atomic force microscopy, and magnetometry.

[22] J. Lewis, D. Schwarzenbach, and H. D. Flack, Acta Crystallogr. Sect. A **38**, 733 (1982).

[23] L. W. Finger, R. M. Hazen, and A. M. Hofmeister, Phys. Chem. Miner. **13**, 215 (1986).

[24] C. J. Fennie and K. M. Rabe, Phys. Rev. B **72**, 100103 (2005).

[25] H. Wang, I. V. Solovyev, W. Wang, X. Wang, P. P. J. Ryan, D. J. Keavney, J.-W. Kim, T. Z. Ward, L. Zhu, J. Shen, X. M. Cheng, L. He, X. Xu, X. Wu, J. Shen, L. He, X. Xu, and X. Wu, Phys. Rev. B **90**, 14436 (2014).

[26] H. Das, A. L. Wysocki, Y. Geng, W. Wu, and C. J. Fennie, Nat Commun **5**, 2998 (2014).

[27] D. T. Margulies, F. T. Parker, M. L. Rudee, F. E. Spada, J. N. Chapman, P. R. Aitchison, and a E. Berkowitz, Phys. Rev. Lett. **79**, 5162 (1997).

[28] W. Eerenstein, T. T. M. Palstra, T. Hibma, and S. Celotto, Phys. Rev. B **66**, 201101 (2002).

[29] X. H. Liu, A. D. Rata, C. F. Chang, A. C. Komarek, and L. H. Tjeng, Phys. Rev. B **90**, 125142 (2014).

[30] E. Della Torre, *Magnetic Hysteresis* (IEEE Press, New York, 1999).

[31] J. Stöhr, J. Magn. Magn. Mater. **200**, 470 (1999).

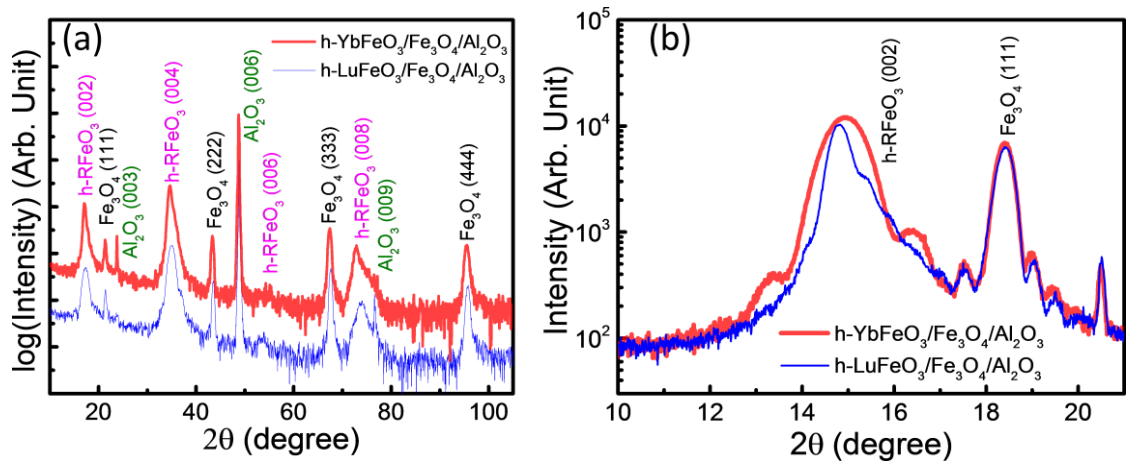

**Figure 1**. X-ray diffraction of the h-RFeO$_3$/Fe$_3$O$_4$/Al$_2$O$_3$ films. (a) Large range θ-2θ scan using a cobalt K-$\alpha$ source ($\lambda$ = 1.79 Å). (b) Small range θ-2θ scan using a copper K-$\alpha$ source ($\lambda$ = 1.45 Å). The ripples in the diffraction peaks are the Laue oscillations.

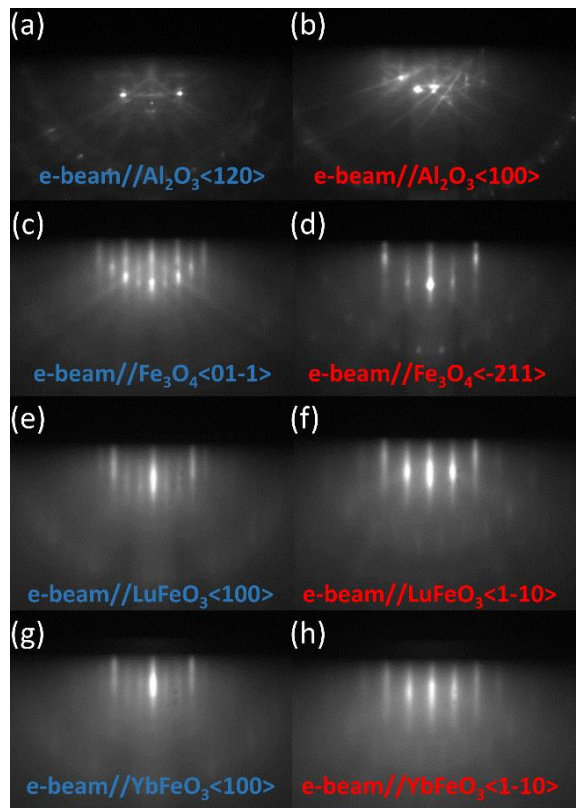

**Figure 2.** RHEED images of different surfaces with two perpendicular directions of incident electron beams relative to the substrate. In (a), (c), (e), and (g), the orientation of the substrate is fixed so that the electron beam is parallel to $Al_2O_3$ <120>. In (b), (d), (f), and (h), the orientation of the substrate is fixed so that the electron beam is parallel to $Al_2O_3$ <100>. The alignment between the electron beams and the films lattices are also indicated.

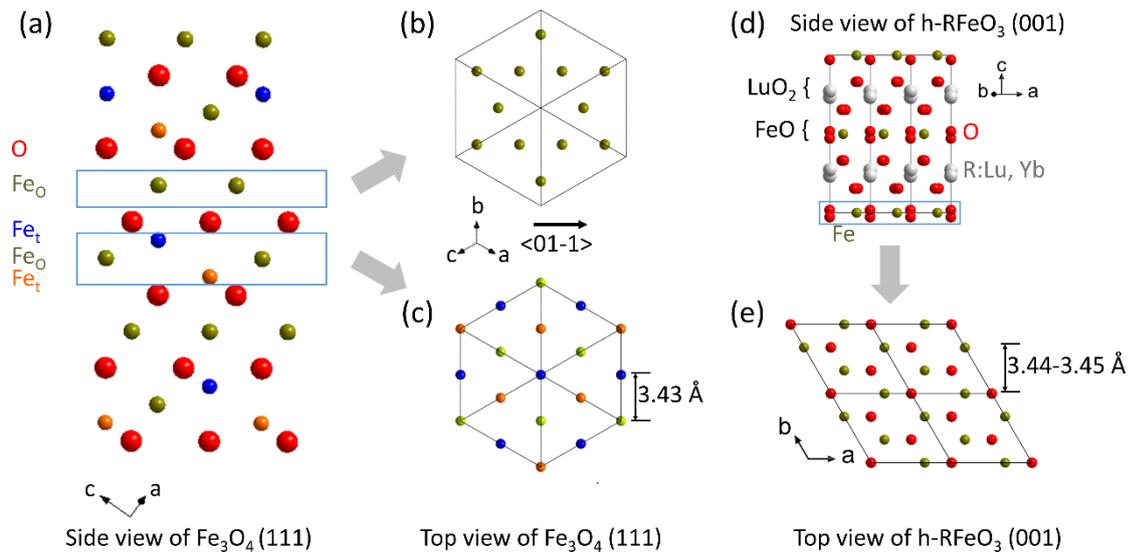

**Figure 3.** Structural model at the h-RFeO$_3$ (001)/Fe$_3$O$_4$ (111) interfaces. (a) Side view of the crystal structure of the Fe$_3$O$_4$ (111) film; the two kinds of Fe layers that are parallel to the Fe$_3$O$_4$ (111) plane are indicated by the boxes. Fe$_o$ and Fe$_t$ are Fe sites in oxygen octahedral and oxygen tetrahedral environments respectively. (b) and (c) are the top views of the two kinds of Fe layers indicated in (a). (d) Side view of the crystal structure of the h-RFeO$_3$ (001) film, where the FeO layer is indicated by the box. (e) Top view of the FeO layer in h-RFeO$_3$ (001) indicated in (d). (b-e) are in the same scale.

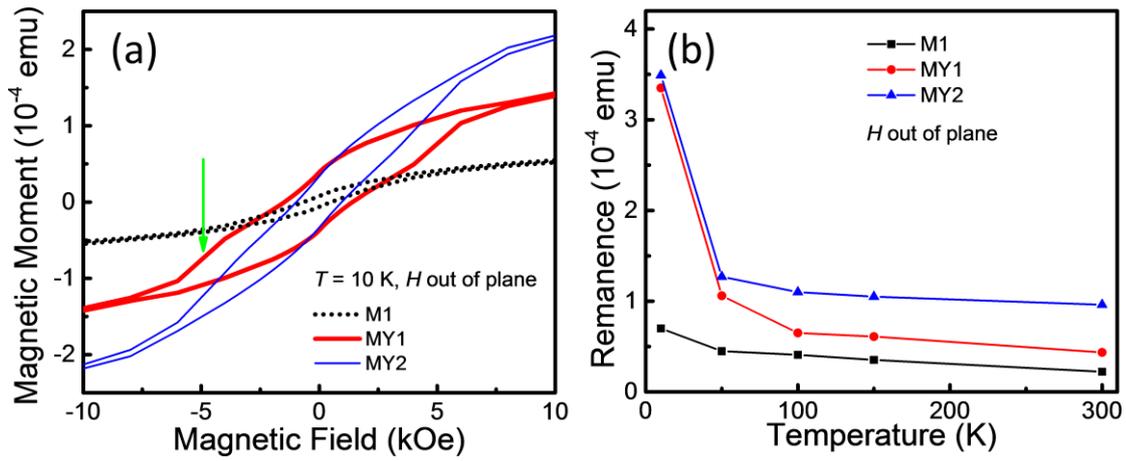

**Figure 4**. (a) Magnetic hysteresis loops for three film samples: $Fe_3O_4$ (8.5 nm)/$Al_2O_3$ (M1), h-$YbFeO_3$ (25 nm)/$Fe_3O_4$ (11 nm)/$Al_2O_3$ (MY1), h-$YbFeO_3$ (20 nm)/$Fe_3O_4$ (21 nm)/$Al_2O_3$ (MY2), measured at 10 K with magnetic field along the out-of-plane direction. (b) The magnetic remanence of the three film samples as a function of temperature.

# Effect of interface on epitaxy and magnetism in h-RFeO₃/Fe₃O₄/Al₂O₃ films (R=Lu, Yb): Supplementary materials

## 1. Atomic force microscopy

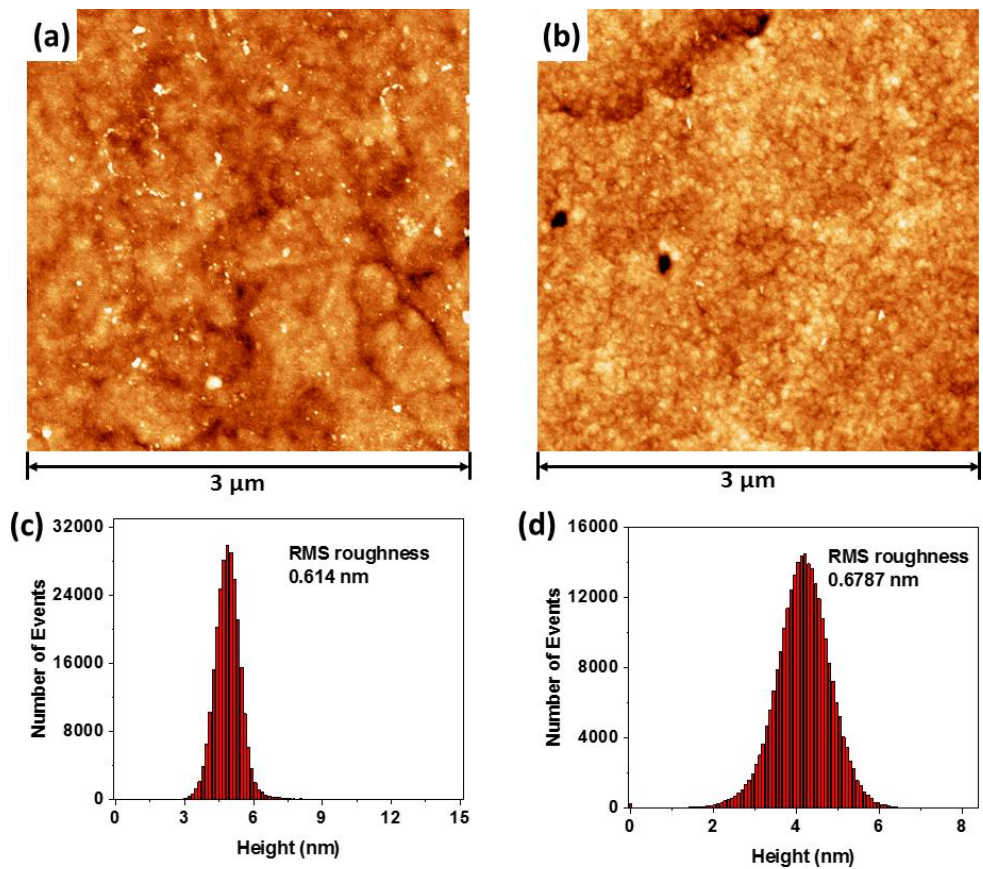

**Figure S1**. Atomic force microscopy on h-LuFeO₃ (a) and (c), and on h-YbFeO₃ (b) and (d).

## 2. X-ray Diffraction

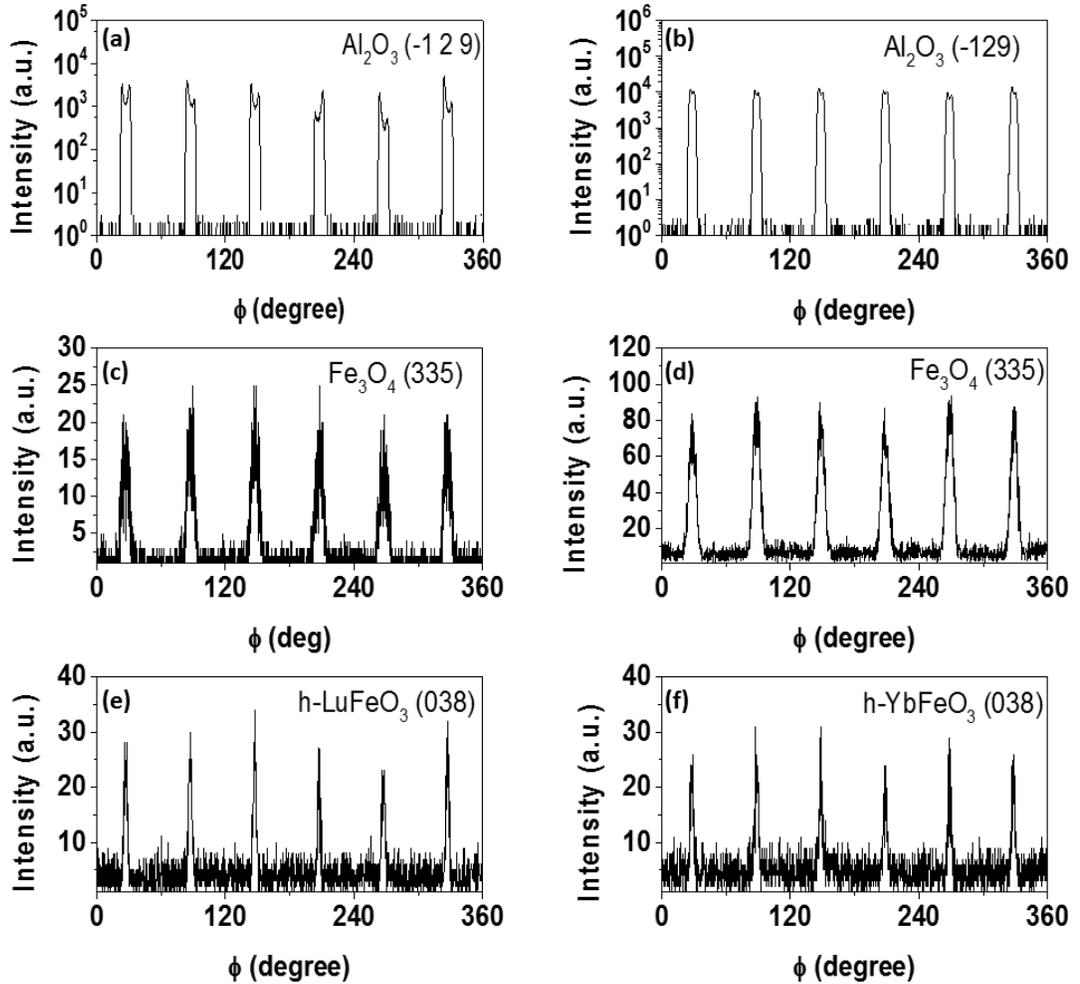

**Figure S2**. φ scan of an h-LuFeO$_3$ (001)/Fe$_3$O$_4$ (111)/Al$_2$O$_3$ (001) film: left column (a), (c), and (e), and an h-YbFeO$_3$ (001)/Fe$_3$O$_4$ (111)/Al$_2$O$_3$ (001) film: right column (b), (d), and (f).

According to the φ scan of x-ray diffraction, the in-plane epitaxial relations are h-RFeO$_3$ (010)//Fe$_3$O$_4$ (110)//Al$_2$O$_3$ (-120). If these relations are converted to crystal axis in real space, one has h-RFeO$_3$ <100> //Fe$_3$O$_4$ <01-1>//Al$_2$O$_3$ <120>.

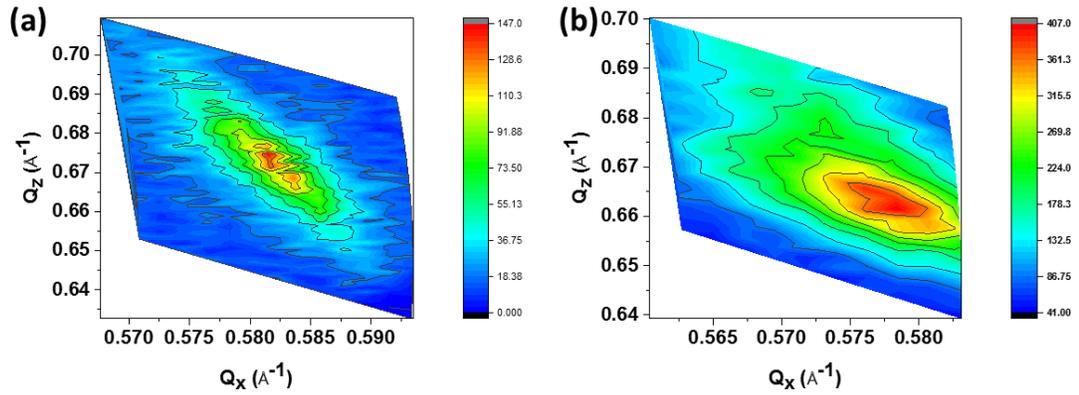

**Figure S3**. Reciprocal space mapping (RSM) using x-ray diffraction on the h-RFeO$_3$ (038) peak of an h-LuFeO$_3$ (001)/Fe$_3$O$_4$ (111)/Al$_2$O$_3$ (001) film and an h-YbFeO$_3$ (001)/Fe$_3$O$_4$ (111)/Al$_2$O$_3$ (001) film.

According to the reciprocal space mapping, the peak position for h-LuFeO$_3$ (038) is $Q_z$ =0.6716 Å$^{-1}$, $Q_x$ =0.5809 Å$^{-1}$, corresponding to the lattice constants a = 5.963 Å; c =11.91 Å; the peak position for the h-YbFeO$_3$ (038) is $Q_z$ = 0.6627 Å$^{-1}$ and $Q_x$ = 0.5754 Å$^{-1}$, corresponding to the lattice constants a = 6.020 Å and c = 12.08 Å.

## 3. Magnetometry measurements and discussions

The magnetic field dependence of magnetization *M(H)* is a normal means to study the magnetic anisotropy. In principle, the magnetic anisotropy can be extracted by comparing the high field M(H) relations for different magnetic field directions. In addition, derived from the magnetic anisotropy, is the hysteresis in *M(H)* (and the corresponding remanence and coercivity) when the magnetic domain motion is pinned. Normally, a high magnetic coercivity can be obtained in a material of high magnetic anisotropy and with the magnetic field along the easy axis.

For thin films, there are magneto-crystalline anisotropy as well as the shape anisotropy. The shape anisotropy is generated by the anisotropy of the depolarization factor in a film due to its quasi two-dimensional shape. While the direction of the magneto-crystalline anisotropy depends on the specific crystal structure, for the shape anisotropy of a thin film, the hard axis is always along the out-of-plane direction.

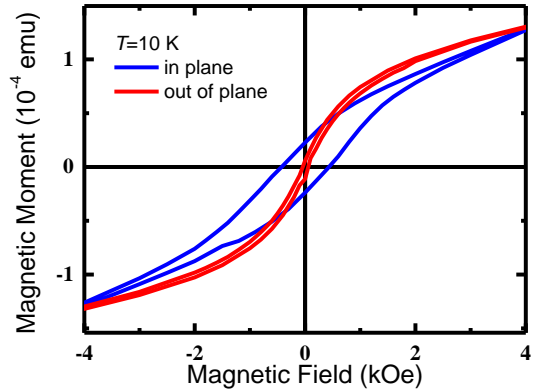

**Figure S4**. Magnetic hysteresis loops for a $Fe_3O_4$ (8.5 nm) / $Al_2O_3$ (M1) film, measured at 10 K with magnetic field along the in-plane and out-of-plane directions.

Therefore, it is not trivial to extract the magnetic anisotropy information from the *M(H)* in the $Fe_3O_4$ and h-$YbFeO_3$ films. In $Fe_3O_4$ films, the magneto-crystalline anisotropy is often dominated by the anisotropy created by one type of defect called anti-phase boundary. [1–3] This kind of anisotropy exists in all crystalline directions, contribute little to the remanence and coercivity, but results in none-zero slope in M(H) up to 70 kOe and. [1–3] As shown in Fig. S4, the magnetization of a $Fe_3O_4$ (8.5 nm)/$Al_2O_3$ (M1) film was measured at 10 K with magnetic field along the in-plane and out-of-plane directions. According to the remanence and coercivity, the out-of-plane direction is the easy axis, which is consistent with the shape anisotropy. At the same time, the high field magnetizations are far from being saturated, which is consistent with the anisotropy generated by the anti-phase boundaries.

The $M(H)$ relation is also special in h-YbFeO$_3$ films. The ferromagnetic moments in the hexagonal ferrites are parasitic to its antiferromagnetic order, in which all the Fe moments lie in the basal plane. [4] The ferromagnetic moments in h-YbFeO$_3$ films comes from the canting of the Fe moments toward the out-of-plane direction and the further polarization of the (paramagnetic) Yb moments. The net in-plane moment from Fe is symmetry forbidden. Therefore, in h-YbFeO$_3$ films, the shape anisotropy is not expected to play a role. On the other hand, the high field M(H) is dominated by the paramagnetic response of Yb, which contributes a non-zero slope up to a very high field. [5]

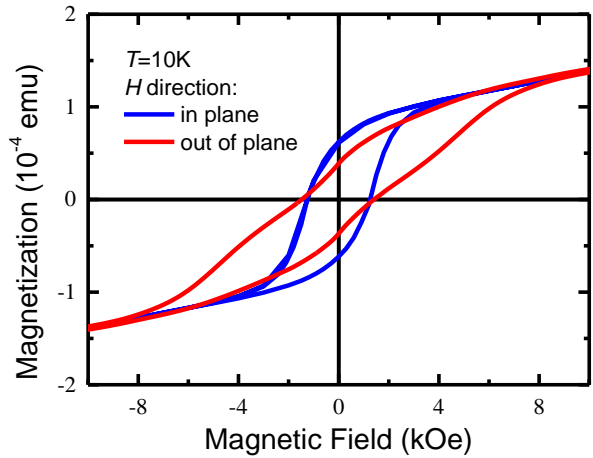

**Figure S5**. Magnetic hysteresis loops for h-YbFeO$_3$ (25 nm) / Fe$_3$O$_4$ (11 nm) / Al$_2$O$_3$ (MY1), measured at 10 K with magnetic field along the in-plane and the out-of-plane directions.

Keeping in mind the discussion above, we are ready to examine the $M(H)$ measurement of the h-YbFeO$_3$ (001)/Fe$_3$O$_4$ (111)/Al$_2$O$_3$ (001) films. As shown in Fig. S5, the magnetization of the h-YbFeO$_3$ (25 nm)/Fe$_3$O$_4$ (11 nm)/Al$_2$O$_3$ (MY1) film was measured for the magnetic field along the in-plane and compared with that for the magnetic field along the out-of-plane direction. In the in-plane magnetic field, the M(H) is similar to that of the M1 film shown in Fig. S4. In the out-of-plane magnetic field, the M(H) loop show two features. One corresponds to the reversal of the ferromagnetic moments in the h-YbFeO$_3$, and the other corresponds to the magnetization of the Fe$_3$O$_4$ layer (see Fig. S4). Therefore, it appears that the anisotropy of the h-YbFeO$_3$ (001) and Fe$_3$O$_4$ (111) are preserved in the h-YbFeO$_3$ (001)/Fe$_3$O$_4$ (111)/Al$_2$O$_3$(001) film: the out-of-plane direction is the easy axis for the h-YbFeO$_3$ (001) and the hard axis for the Fe$_3$O$_4$ (111).

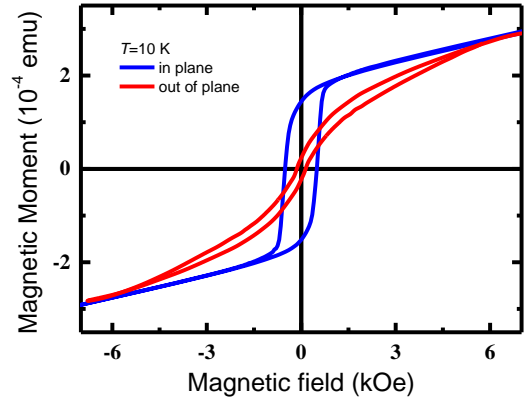

**Figure S6**. Magnetic hysteresis loops for h-LuFeO$_3$ (001)/ Fe$_3$O$_4$ (111)/ Al$_2$O$_3$ (001), measured at 10 K with magnetic field along the in-plane and the out-of-plane directions.

For the h-LuFeO$_3$ (001)/Fe$_3$O$_4$ (111)/Al$_2$O$_3$ (001) films, since the net moment from h-LuFeO$_3$ is very small ($\approx 0.02$ $\mu_B$/f.u.) [6,7], the magnetometry measurements basically characterizes the magnetic properties of the Fe$_3$O$_4$ (111) layer. As shown in Fig. S6, the magnetization of an h-LuFeO$_3$ (001)/Fe$_3$O$_4$ (111) Al$_2$O$_3$ (001) film was measured at 10 K with magnetic field along the in-plane and the out-of-plane directions. Both the *M*(*H*) of the in-plane and out-plane directions look similar to those of the Fe$_3$O$_4$ (111) film in Fig. S4, with a clear indication of a hard axis along the out-of-plane direction.

**Reference**

[1]   D. T. Margulies, F. T. Parker, M. L. Rudee, F. E. Spada, J. N. Chapman, P. R. Aitchison, and a E. Berkowitz, Phys. Rev. Lett. **79**, 5162 (1997).

[2]   W. Eerenstein, T. T. M. Palstra, T. Hibma, and S. Celotto, Phys. Rev. B **66**, 201101 (2002).

[3]   X. H. Liu, A. D. Rata, C. F. Chang, A. C. Komarek, and L. H. Tjeng, Phys. Rev. B **90**, 125142 (2014).

[4]   W. Wang, J. Zhao, W. Wang, Z. Gai, N. Balke, M. Chi, H. N. Lee, W. Tian, L. Zhu, X. Cheng, D. J. Keavney, J. Yi, T. Z. Ward, P. C. Snijders, H. M. Christen, W. Wu, J. Shen, and X. Xu, Phys. Rev. Lett. **110**, 237601 (2013).

[5]   Y. K. Jeong, J. Lee, S. Ahn, S.-W. Song, H. M. Jang, H. Choi, and J. F. Scott, J. Am. Chem. Soc. **134**, 1450 (2012).

[6]   J. A. Moyer, R. Misra, J. A. Mundy, C. M. Brooks, J. T. Heron, D. A. Muller, D. G. Schlom, and P. Schiffer, APL Mater. **2**, 12106 (2014).

[7]   X. Xu and W. Wang, Mod. Phys. Lett. B **28**, 1430008 (2014).